\documentclass[10pt, conference, compsocconf]{IEEEtran}
\ifCLASSINFOpdf
\else
\fi
\usepackage{booktabs} 
\usepackage{comment}
\usepackage{enumerate}
\usepackage{color}
\usepackage{graphicx}
\usepackage{subfigure}
\usepackage{amsfonts}
\usepackage{amsmath}
\usepackage{multirow, makecell, caption}

\begin{document}
%
\title{How to Become Instagram Famous: Post Popularity Prediction with Dual-Attention}


\author{\IEEEauthorblockN{Zhongping Zhang}
\IEEEauthorblockA{Electrical and Computer Engineering\\
University of Rochester\\
Rochester, NY 14627\\
zhongping.vista@gmail.com}
\and
\IEEEauthorblockN{Tianlang Chen}
\IEEEauthorblockA{Department of Computer Science\\
University of Rochester\\
Rochester, NY 14627\\
tchen45@cs.rochester.edu}
\and
\IEEEauthorblockN{Zheng Zhou}
\IEEEauthorblockA{Department of Electrical Engineering
\\University at Buffalo\\
Buffalo, NY 14260\\
zzhou32@buffalo.edu}
\and
\IEEEauthorblockN{Jiaxin Li}
\IEEEauthorblockA{Department of Electrical Engineering\\
Harbin Institute of Technology\\
Harbin, China, 150000\\
lijiaxinpp93@yeah.net}
\and
\IEEEauthorblockN{Jiebo Luo}
\IEEEauthorblockA{Department of Computer Science\\
University of Rochester\\
Rochester, NY 14627\\
jluo@cs.rochester.edu}
}


%


\maketitle

\begin{abstract}
With a growing number of social apps, people have become increasingly willing to share their everyday photos and events on social media platforms, such as Facebook, Instagram, and WeChat. In social media data mining, post popularity prediction has received much attention from both data scientists and psychologists. Existing research focuses more on exploring the post popularity on a population of users and including comprehensive factors such as temporal information, user connections, number of comments, and so on. However, these frameworks are not suitable for guiding a specific user to make a popular post because the attributes of this user are fixed. Therefore, previous frameworks can only answer the question ``whether a post is popular'' rather than ``how to become famous by popular posts''. In this paper, we aim at predicting the popularity of a post for a specific user and mining the patterns behind the popularity.
To this end, we first collect data from Instagram. We then design a method to figure out the user environment, representing the content that a specific user is very likely to post.  Based on the relevant data, we devise a novel dual-attention model to incorporate image, caption, and user environment. The dual-attention model basically consists of two parts, explicit attention for image-caption pairs and implicit attention for user environment. A hierarchical structure is devised to concatenate the explicit attention part and implicit attention part.
We conduct a series of experiments to validate the effectiveness of our model and investigate the factors that can influence the popularity. The classification results show that our model outperforms the baselines, and a statistical analysis identifies what kind of pictures or captions can help the user achieve a relatively high ``likes'' number.

\end{abstract}

\begin{IEEEkeywords}
Popularity Prediction, Dual-Attention Model, Instagram, Social Media Data Mining

\end{IEEEkeywords}

%
\IEEEpeerreviewmaketitle

\section{Introduction}

\begin{figure}[h]
\centering
\includegraphics[width=3.5in]{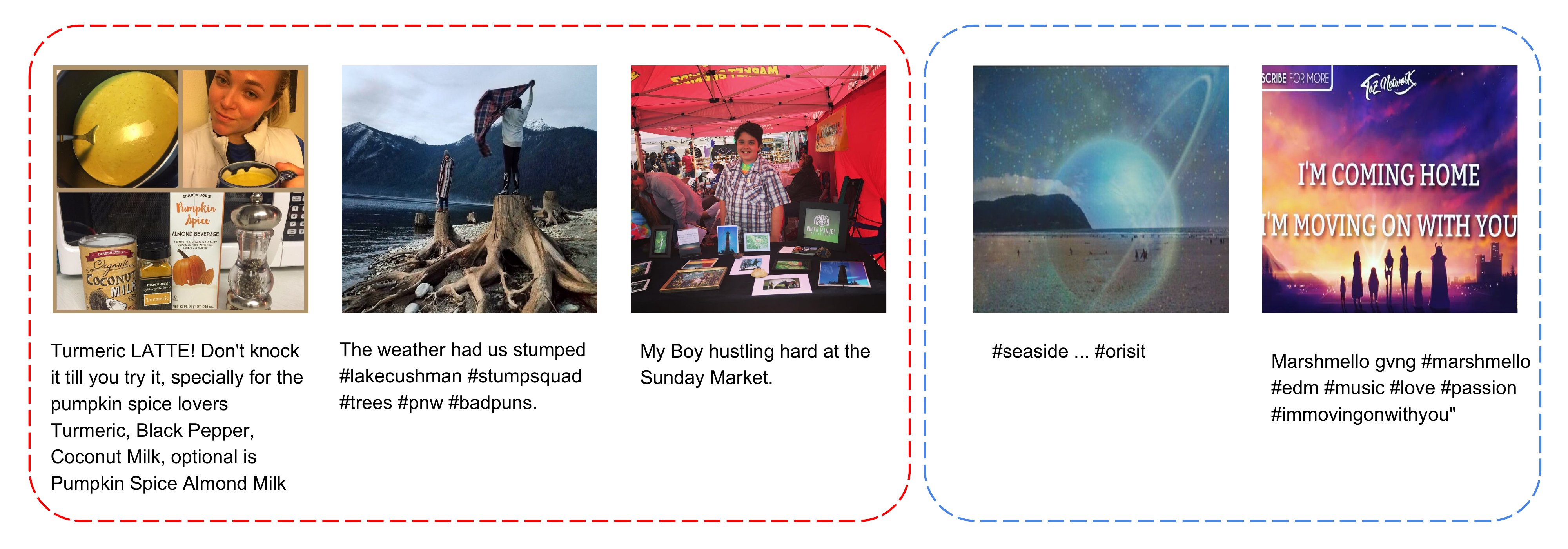}
\caption{ Example posts in Instagram. Posts in the red dotted box are popular. Posts in the blue dotted box are unpopular.}
\label{Fig: img_exp}
\end{figure}

\begin{figure*}
\centering
\vspace{0.5cm}
\includegraphics[width=6.4in]{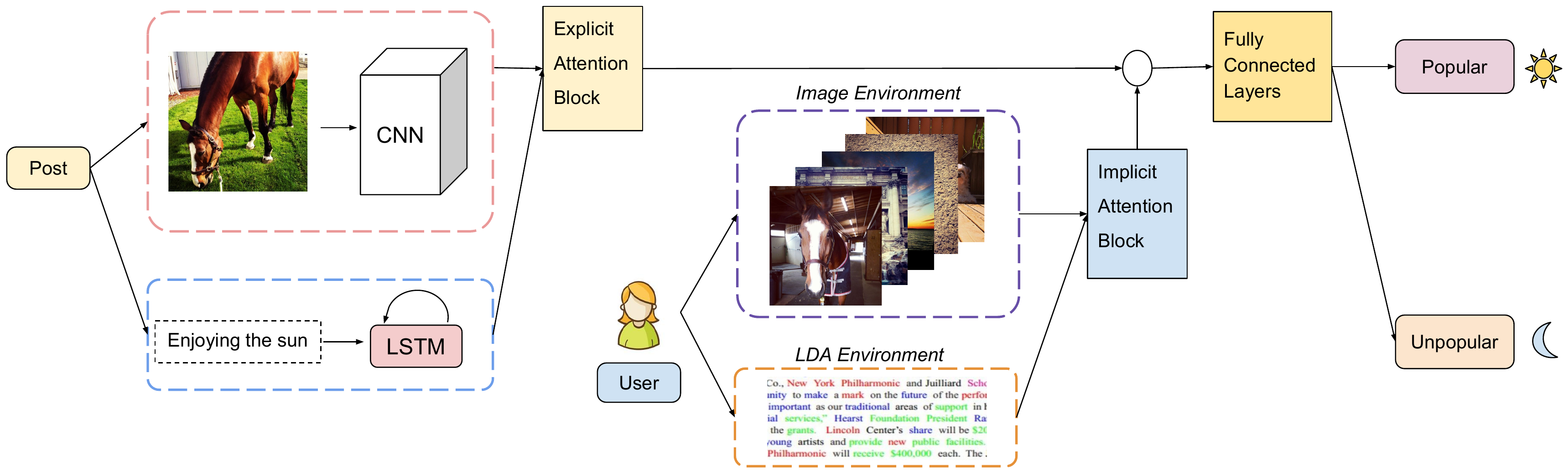}
\caption{ Overview of the dual-attention model.}
\label{Fig: overview}
\end{figure*}
Social media platforms provide their users with a good opportunity to share daily lives, emotions, and so on. Driven by data, post popularity prediction has been focused and studied widely in recent years. Researchers are able to build powerful models to predict post popularity from various aspects, such as image \cite{gelli2015image}\cite{totti2014impact}, textual content \cite{hong2011predicting}\cite{szabo2010predicting}, time series \cite{wu2016time}, sentiment \cite{bae2012sentiment} or even brand information \cite{de2012popularity}. These frameworks always measure the popularity of a post from the view of the whole social media platform. For example, the authors in \cite{gelli2015image} and \cite{hong2011predicting} respectively use the number of views and the forwarding number as measurements. Though these indexes can measure popularity from a big picture level, they ignore the diversity of users. For instance, a post with 100 $\sim$ 200 views is popular for a new user. However, a post with 10000 $\sim$ 20000 views still might not be popular for a famous star on the same platform. Under this circumstance, the above measurements are not able to reflect the post popularity for a particular user. A big shot always obtains high popularity scores while a green hand tends to be assigned a low popularity score. In practice, post popularity prediction for a particular user is significant for both companies and their customers. Companies are able to maximize their influence and provide more compelling content for their users. Customers who want to become more attractive can evaluate their posts before they upload them. Motivated by these benefits, we raise a new target in this paper: post popularity prediction for a specific user. To solve the problem, we develop a system which takes the user's images and captions as input and then generates the popularity prediction result. To make our discussion more straightforward, we illustrate several examples in Figure \ref{Fig: img_exp}.

We formulate our task as a binary classification problem to classify whether a post is popular for a particular user. In this paper, a novel dual-attention model is proposed to predict the result. Concretely, the dual-attention model includes two parts: explicit attention model and implicit attention model. These two models take different levels of information as input, and then they are concatenated by a hierarchical structure. Specifically speaking, the explicit attention model is designed to generate attention weights for captions. We modify a co-attention model \cite{lu2016hierarchical} to make it more suitable for Instagram image-caption pairs. Implicit attention model is applied to incorporate user environment. The user environment includes image environment and topic environment. Since environment does not have any explicit meaning for different positions, an implicit attention model without explicit attention mechanism is deployed here. We use a hierarchical structure to connect the explicit attention model and the implicit attention model. The structure is designed because the user environment contains higher level information than image or caption alone. Figure \ref{Fig: overview} demonstrates the framework of our model.

In this study, we collect our data from Instagram, a platform where people can share their pictures and emotions. The dataset contains 441 users and 60,785 image-caption pairs. Based on image-caption pairs, we extract the user environment and feed it into the proposed implicit attention model. Our target is not limited to predict the popularity of a post, but also explore the correlations between image, caption, and popularity. A series of experiments is therefore performed to evaluate our model and reveal the correlations. 

The main contributions of our work are:
\begin{list}{$\bullet$}
{ \setlength{\leftmargin}{0.18cm}}
	\item We introduce a framework to address the problem of post popularity prediction for a specific user.
	\item We propose a method to calculate user environment. Compared with image and caption, the user environment is a higher level information. It can provide the model with a more comprehensive understanding of users, thus can further improve the performance of the model.
    \item We develop a novel dual-attention model to predict whether a post is popular. The dual-attention model consists of two parts, an explicit attention part for image-caption pairs and an implicit attention part for user environment. 
    \item We perform two levels of experiments on the Instagram dataset. First, we present the classification results to demonstrate the effectiveness of our framework. Next, we explore the factors which can influence the user's post popularity.
\end{list}

\section{Related Work}
Our work is mainly related to user trait pattern, popularity prediction, and attention model. In this section, we will respectively discuss the related work from these three aspects.

\subsection{User Trait Pattern}
With recent advances in social media data mining, exploring user trait behind data has become a popular research topic. Image information is extensively used to provide valuable cues for identifying user attributes. In \cite{you2014eyes}, the authors use images posted on various social networks to infer the user gender. More recently, Dhir et al. \cite{dhir2016age} attempt to predict age and gender from selfie-related behavior. Besides, there are several efforts to explore the inner traits of a person, such as personality \cite{liu2016analyzing}, interest \cite{you2016picture}, etc. Topic information is another important source for user trait prediction. One of the most influential papers in topic model is Latent Dirichlet Allocation (LDA) \cite{blei2003latent}. LDA is a generative probabilistic model which is widely used to extract topics from unlabeled documents. Based on the topic model and text messages, many interesting studies such as \cite{hu2016language}\cite{hamidian2015rumor} are presented in recent years. 

In our paper, we apply LDA topic model to construct user topic environment. We will discuss more details in Section \ref{model:user_env}. 

\subsection{Attention Model}
Attention model is first applied in English-French translation task \cite{kalchbrenner2013recurrent}. Motivated by the success in language translation, many researchers apply attention models in image captioning \cite{xu2015show}\cite{mnih2014recurrent}\cite{qi2018stagnet}. To get a better representation of image and caption, co-attention model \cite{xiong2016dynamic}\cite{lu2016hierarchical} are proposed. Since co-attention mechanism considers both the image and caption, it can generate affinity matrix which includes not only the spatial attention weights for image but also the text attention weights for caption. Considering the attention weights can be extracted from a certain layer of the model, this kind of attention mechanism can be defined as explicit attention mechanism. On the other hand, Kim et al. \cite{kim2016multimodal} propose an implicit attention without explicit attention parameters. They apply the structural similarity with residual learning to avoid the attention parameters, but still effectively learns the joint representation from vision and natural language.

\subsection{Popularity Prediction}
The technological and economic importance of popularity prediction motivate many researchers to notice this area \cite{hong2011predicting}\cite{szabo2010predicting}
\cite{niu2012predicting}\cite{mcparlane2014nobody}
\cite{wu2016time}\cite{wu2016unfolding}
\cite{bae2012sentiment}\cite{de2012popularity}\cite{mazloom2016multimodal}
\cite{qi2017online}. 
Image is the main research direction of popularity prediction. The authors in \cite{niu2012predicting}\cite{mcparlane2014nobody}\cite{khosla2014makes}\cite{gelli2015image}\cite{totti2014impact} explore image popularity based on information extracted from image, like objects, image metadata and so on. Compared with image, text is another hot research area. In \cite{hong2011predicting}\cite{szabo2010predicting}, the authors predict on-line message popularity by analyzing textual information. Besides image and text, some novel information such as time \cite{wu2016time}\cite{wu2016unfolding}, sentiment \cite{bae2012sentiment} or even brand \cite{de2012popularity}\cite{mazloom2016multimodal} are also considered to guide the prediction.

As we discussed above, previous work on popularity prediction always focuses on the big picture level and ignores the diversity of users. Furthermore, many of these frameworks attempt to incorporate extra information like friend links, user contacts, sentiments, and user tags to improve the accuracy of prediction. However, the above information is not always available. In this paper, we attempt to develop a system which can predict post popularity for a particular user based only on image-caption pairs.

\section{DUAL-ATTENTION MODEL}
In this section, we introduce a dual-attention model by five steps. Firstly, we start by listing some important notations to avoid ambiguity. Secondly, we present the explicit attention model in Section \ref{model:explicit}. Then, we introduce the user environment and describe the details on how to calculate it in Section \ref{model:user_env}. After that, the implicit attention model is proposed in Section \ref{model:implicit}. Finally, the overall structure of the dual-attention model is described.

\subsection{Notations}
\label{model:notations}
To ease understanding the following parts, here we list several important notations:
\begin{list}{$\bullet$}
{ \setlength{\leftmargin}{0.4cm}}
	\item $Q=\{q_1,..,q_T\}$ denotes a caption with T words, where $q_{t}$ corresponds the onehot vector of the $t$-th word 
	\item $W_{(\cdot)}$ denotes weights of different layers, we omit bias to avoid redundancy
    \item $\sigma_{(\cdot)}$ denotes activation functions of different layers
    \item $\mathbb{F}(\cdot)$ denotes fully connected layer
    \item $\hat{V}$ and $\hat{Q^e}$ denote the attended features of image and caption
    \item $I_e$ and $T_e$ denote image environment and topic environment respectively
    \item $F(I_e,T_e)$ and $H(q,v)$ denote joint residual function and optimal mapping respectively, these two notations are consistent with the definition in \cite{kim2016multimodal}
\end{list}

\begin{figure}
\centering
\includegraphics[width=3in]{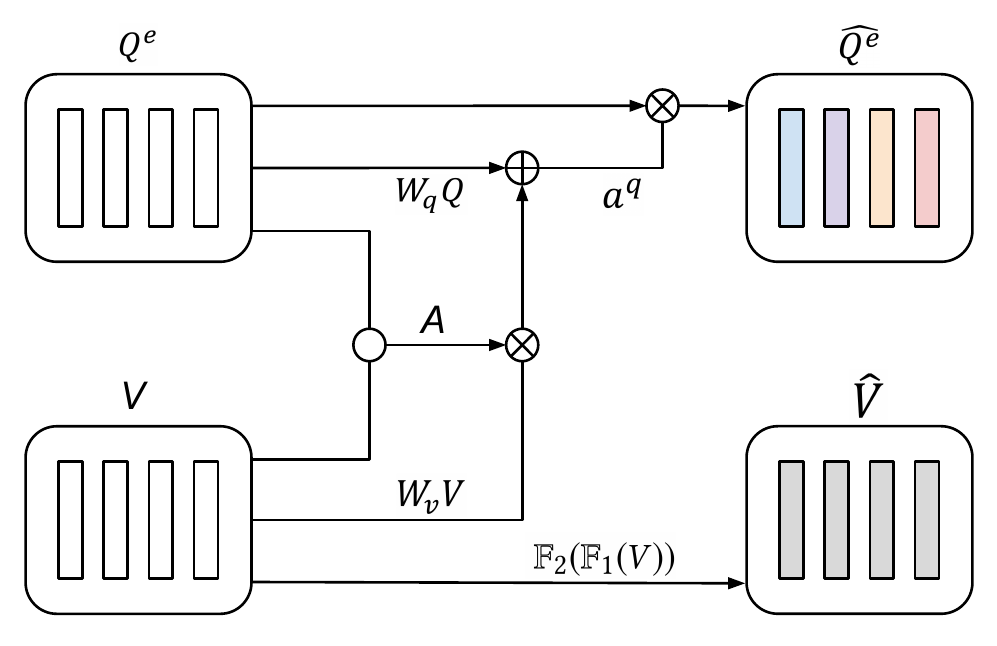}
\caption{Illustration of the explicit attention model.}
\label{Fig: explicit_attn}
\end{figure}

\subsection{Explicit Attention Model}
\label{model:explicit}
Given an image $I$ and its corresponding caption $Q = \{q_1,...,q_T\}$, we first encode them into a feature vector space. ResNet-50 \cite{he2016deep}, which is pre-trained on ImageNet dataset \cite{krizhevsky2012imagenet}, is applied here as the image encoder. We extract the image features $V$ from the last pooling layer, whose dimension is 2048. To get word features $Q^e = \{q^e_1,...,q^e_T\}$, we embed the captions with a word embedding layer followed by a one-layer LSTM.

Co-attention mechanism \cite{nguyen2018improved}\cite{lu2016hierarchical} are commonly used in order to get a better representation of images and words.
However, unlike the traditional visual question-answering (VQA) or image captioning tasks, Instagram images usually do not contain complex spatial information because most pictures uploaded by users are selfies, landscapes, posters and so on. Most of these pictures do not express complicated logical relationship nor different importance among objects. Therefore, we modify the co-attention model based on \cite{lu2016hierarchical} to make it more suitable for our popularity prediction task. 

Concretely, the attention model starts with calculating the affinity matrix $A$ between image $V\in R^{d_1}$ and caption $Q^e \in R^{d_2 \times T}$ representations
\begin{equation}
\begin{aligned}
A = tanh( (Q^e)^TW_aV)
\end{aligned}
\end{equation}
where $W_a\in R^{d_2\times d_1}$ is the learning matrix. Elements in $A\in R^{T\times 1}$ are affinity scores between image and each word. According to affinity matrix $A$, we can further calculate the attention weights $a^q$ via the following equations
\begin{equation}
\begin{split}
H^q &= tanh(W_qQ + A\cdot W_vV)\\
a^q &= softmax(W_{h}H^q)\\
\end{split}
\end{equation}
where $W_q\in R^{k\times d_2}$, $W_v\in R^{k\times d_1}$ and $W_h\in R^{k\times 1}$ are all learning matrices for the explicit attention model. $k$ represents the last dimension of $H^q$, and here we manually set $k$ as 128. Note that though we only obtain the attention weights for captions, image information is still involved during the process of calculating $a^q$.

Finally, the new representations of image and caption are
\begin{equation}
\begin{split}
& \hat V = \mathbb{F}_2(\mathbb{F}_1(V))\\
& \hat{Q^e} = \sum^{T}_{t=1}a^q_t q^e_t
\end{split}
\end{equation}
$\mathbb{F}(\cdot)$ means fully connected layer. We apply $\mathbb{F}(\cdot)$ to map the image features $V$ into the same dimension as caption features $\hat{Q^e}$.

We demonstrate the structure of our attention model in Figure \ref{Fig: explicit_attn}. Since the parameters of attention weights are explicitly propagated in this model, this model is named as explicit attention model to distinguish from the implicit attention model in Section \ref{model:implicit}.

\subsection{User Environment}
\label{model:user_env}
In most cases, post popularity is not only influenced by its corresponding image and caption but also rely on the user who makes the post. For example, a person whose picture wall is full of landscapes, upload a selfie one day. The selfie is very likely to get a high number of ``likes''. On the contrary, if the user is selfie-addicted and uploads selfie every day, a new selfie is less likely to be popular because his friends have got used to it. Motivated by this circumstance, we introduce the concept of user environment to further improve our model. 

We utilize user environment to indicate the content that the user is very likely to post. Therefore, we introduce the average value of user features to represent the environment. Image environment $I_v$ is directly defined as the mean value of the deep-level image features $V$. With respect to topic environment $T_e$, Latent Dirichlet Allocation (LDA) \cite{blei2003latent} is applied to assign a topic feature to each caption. LDA is a generative statistical model of corpus. It assumes that documents have several random latent topics, and each topic can be characterized by a distribution over words. In our system, we set the number of topics as 400 which means the LDA feature of each caption is a 400 dimension vector. Similar with image features, we also use the mean value of LDA features to represent topic environment.

\subsection{Implicit Attention Model}
\label{model:implicit} 
Given image environment $I_e$ and topic environment $T_e$, the most direct method to incorporate them is using fully connected layers followed by a concatenation layer. Considering the structure of the fully connected layer cannot highlight important positions on the environment features, we apply an attention model as an alternative choice in the environment-encoding process. Different from common image and text fusion target, it is a challenging task to explicitly express what the elements of environment features stand for. Therefore we devise an implicit attention model motivated by \cite{kim2016multimodal}. (Consistent with \cite{kim2016multimodal}, we still use $F$ to denote joint residual function, and use $H$ to denote optimal mapping.)

\begin{figure}
\centering
\includegraphics[width=3in]{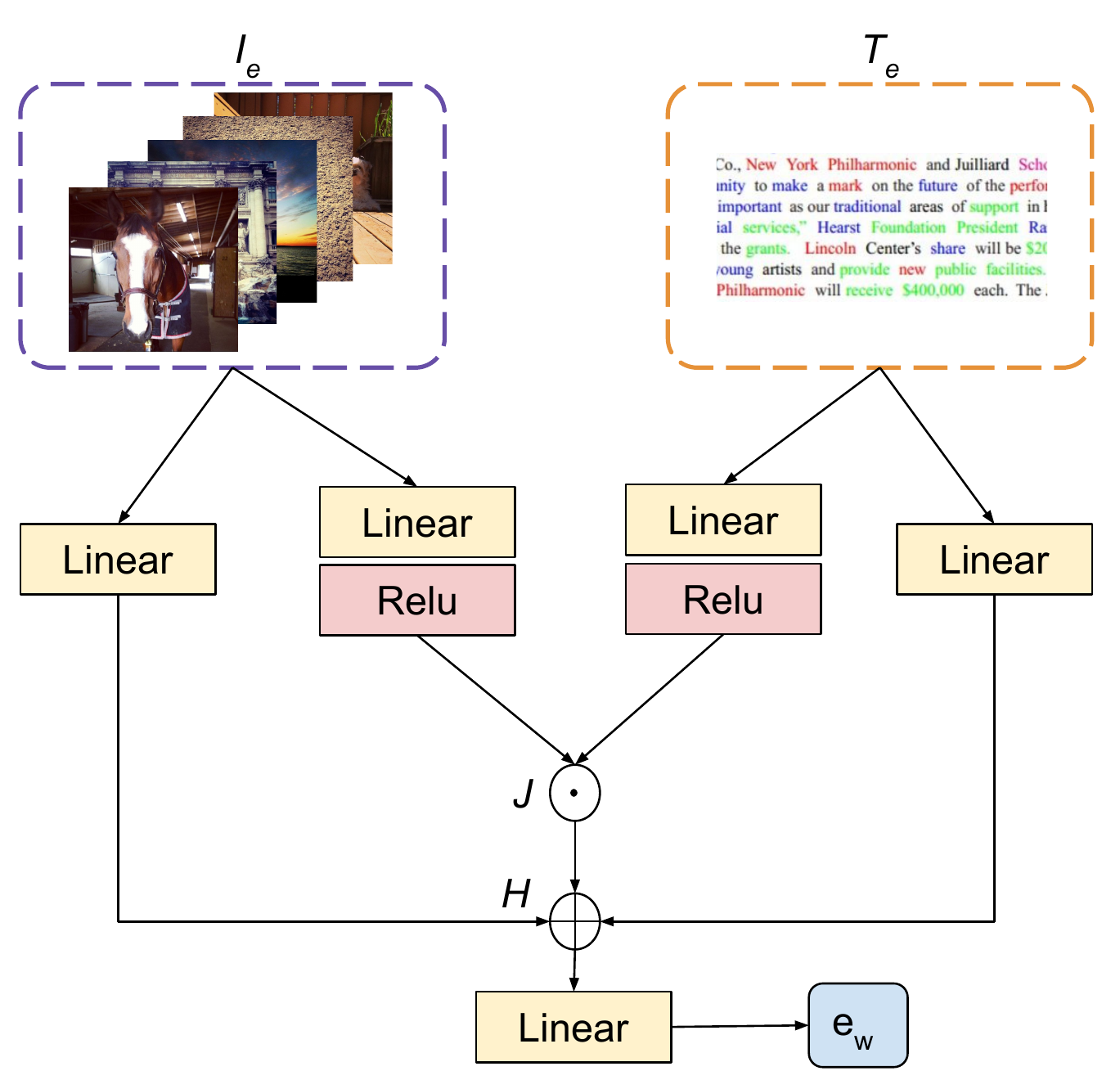}
\caption{Structure of the implicit attention model.}
\label{Fig: implicit_attn}
\end{figure}

Different from explicit attention models, the attention parameters of implicit attention model are hidden in the element-wise multiplication layer. We present our implicit attention model in Figure \ref{Fig: implicit_attn}. As the figure shows, user environment variables $I_e$ and $T_e$ are fed into a fully connected layer respectively in our model. The joint residual function is given by
\begin{equation}
\begin{split}
F(I_e,T_e) = \sigma(W_i I_e) \odot \sigma(W_t T_e)
\end{split}
\end{equation}
where $\sigma$ is $Relu$ and $\odot$ is element-wise multiplication. $W_i$ and $W_t$ is used for encoding $I_e$, $T_e$.

Given the joint residual function, optimal mapping $H(q,v)$ is predicted by
\begin{equation}
\begin{split}
H(q,v) = W_{i2} I_e + W_{t2} T_e + F(I_e,T_e)
\end{split}
\end{equation}
$W_{i2}$ and $W_{t2}$ are shortcut for environment and both of them are encoded into the same feature dimension with $H(q,v)$. After calculating $H(q,v)$, the final representation of environment $e_w$ is obtained by $\mathbb{F}(H(q,v))$.

\subsection{Overall Structure for Predicting Results}
\label{model:overall}
After computing the attended features $\hat V$ and $\hat{Q^e}$, environment representation $e_w$, we predict the final answer through a hierarchical structure as the following equations:
\begin{equation}
\begin{split}
h_w &= (\hat V + \hat{Q^e})\\
O_1 &= \sigma_1(W_1[h_w, e_w])\\
O_2 &= \sigma_2(W_2 O_1)
\end{split}
\end{equation}
where $\sigma_1$ is $Relu$ and $\sigma_2$ is $Sigmoid$ function. $[\cdot] $ indicates concatenation between two tensors. Since we treat the popularity prediction problem as a binary classification task, the loss function is:
\begin{equation}
\begin{split}
\frac{1}{N}\sum_{i=1}^N [y_ilog(\hat y_i)+(1-y_i)log(1-\hat{y_i})]
\end{split}
\end{equation}
where $y_i$ corresponds to ground truth labels and $\hat{y_i}$ indicates predicted labels.

\section{Experiment}
\subsection{Dataset}
There is no public dataset for post popularity prediction. As a result, we collect our data by crawling on Instagram. The dataset we construct contains 441 users and their 60785 image-caption pairs, along with the corresponding number of ``likes''. We choose the number of ``likes'' as the index to measure popularity. In order to consider the popularity for each user, we select top $25\%$ posts (according to the ``likes'' number among each user's posts) as positive samples and bottom $25\%$ posts as negative samples. We randomly select $20\%$ of them as the test set. Besides, we randomly split $10\%$ of training set as the validation set to decide hyper-parameters. In the end, there are 21874 image-caption pairs for training, 2430 image-caption pairs for validation, and 6064 image-caption pairs for testing. The ratio between positive and negative samples is 1:1.

\subsection{Classification Evaluation}
In this section, we conduct classification experiments to evaluate the effectiveness of our proposed model. Considering that our system takes image-caption pairs as input, we choose the following image-caption fusion frameworks as our baselines:
\begin{list}{$\bullet$}
{ \setlength{\leftmargin}{0.4cm}}
	\item \textbf{Single Visual.} The input of Visual model only includes images. We use ResNet-50 which is pretrained on ImageNet to extract image features and feed them into the fully connected layers. 
	\item \textbf{Single Textual} Textual features are first extracted by word embedding layer and one-layer LSTM, then fed into the fully connected layers.
    \item \textbf{Early Fusion.} The image and textual information are concatenated in feature level.
    \item \textbf{Late Fusion.} The image and textual features are fused until the last layer of the model. In another word, the final prediction score can be considered as the average value of visual prediction score and textual prediction score.
    \item \textbf{CCR.} CCR \cite{you2016cross} denotes Cross-modality Consistent Regression Model. It applies KL divergence to measure the consistency between different modality features and concatenated features. 
    \item \textbf{Similarity.} Similar with \cite{fang2015captions}, we use the inner product (cosine similarity) between image and caption as their representation and feed it into the following layers.
\end{list}

In this paragraph, to avoid confusion, we will explain why we do not choose recent popularity prediction frameworks as our baselines. Typical popularity frameworks \cite{gelli2015image}\cite{mazloom2016multimodal}\cite{mazloom2017multimodal} focus on involving more useful information for prediction. Take Mazloom et al. \cite{mazloom2017multimodal} as an example, the authors introduce a three-dimensional tensor which incorporates the user category, item category, and context category. Since the input of their model is a large matrix (the three-dimensional tensor) which already contains obvious and comprehensive information of users, they apply a modified Factorization Machine (FM) to generate prediction results. Similarly, the authors in \cite{gelli2015image} and \cite{mazloom2016multimodal} use Support Vector Regression (SVR) as their prediction model. However, in this paper, the dual-attention model takes the raw image-caption pairs as input. We assume that only the image-caption pairs are available because we aim at predicting the post popularity for particular users. As is known to all, it is meaningless and unfair to compare traditional algorithms like SVR, FM with neural network methods directly on raw data. Therefore, we choose the baselines mentioned above rather than SVR, FM or other traditional algorithms.

For all the baselines and our proposed model, we apply ResNet-50 as the image encoder and one-layer LSTM as the textual encoder. The 2048-dimension image features are extracted from the last pooling layer of ResNet-50. To conduct a fair comparison, we set the dimension of word-embedding features and LSTM hidden state to 512 for all frameworks. During our training process, Adam optimization is used with a learning rate of 0.001 for the first two epochs and with a learning rate of 0.0001 for the following epochs. The size of mini-batch is set to 128.

We demonstrate the quantitative results of our experiments in Table \ref{Experiment: baselines}. The performance of different models is evaluated by four metrics: precision, recall, F-measure, and accuracy. We first compare the baselines with Explicit Attention model. As the table shows, Explicit Attention model can achieve better results under F-measure and accuracy than the other baselines. Although CCR and Late Fusion achieve relatively higher scores in recall and precision respectively, Explicit Attention model obtains a better trade-off under all the metrics. To further improve the model, we include user environment and implicit attention model to construct Dual-attention model. Since the user environment calculation does not rely on any extra information, (our user environments are extracted from users' images and captions), we compare Dual-attention model together with the other models. The results show that Dual-attention model can take one step further based on Explicit Attention model. Almost all metrics of Dual-attention model can exceed $70\%$.

An interesting finding in our experiments is that: Single Textual model performs much better than Single Visual model. For F-measure or Recall, it even achieves better results than some fusion models like Similarity, Late Fusion. Based on this result, we infer that caption information plays a more reliable role than image information on post popularity prediction task.

\begin{table}
\newcommand{\tabincell}[2]{\begin{tabular}{@{}#1@{}}#2\end{tabular}}
\caption{ Comparison of accuracy, precision, recall and F-score }
\label{Experiment: baselines}
\centering
\begin{tabular}{|c|c|c|c|c|}
\hline
 & Precision & Recall & F-measure & Accuracy\\
\hline
Visual & 58.61 & 59.76 & 59.18 & 58.34\\
\hline
Textual & 65.09 & 72.48 & 68.59 & 66.46\\
\hline
Early Fusion & 66.56 & 71.70 & 69.02 & 67.49\\
\hline
Late Fusion & 67.44 & 66.71 & 67.07 & 66.90 \\
\hline
CCR \cite{you2016cross} & 63.90 & 74.99 & 69.01 & 65.96 \\
\hline
Similarity \cite{fang2015captions} & 65.27 & 71.02 & 68.02 & 66.26\\
\hline
\tabincell{c}{Explicit \\Attention(ours)} & 67.25 & 71.64 & 69.38 & 68.05 \\
\hline
\tabincell{c}{Dual-\\Attention(ours)} & \textbf{69.91} & \textbf{75.45} & \textbf{72.58} & \textbf{71.19} \\
\hline
\end{tabular}
\end{table}

\begin{table}
\newcommand{\tabincell}[2]{\begin{tabular}{@{}#1@{}}#2\end{tabular}}
\caption{Comparison of accuracy, precision, recall and F-score in the ablation study }
\label{Experiment: ablation}
\centering
\begin{tabular}{|c|c|c|c|c|}
\hline
 & Precision & Recall & F-measure & Accuracy\\
\hline
Early Fusion &  66.56 & 71.70 & 69.02 & 67.49\\ 
\hline
E-attn &  67.25 & 71.64 & 69.38 & 68.05 \\
\hline
Env & 67.26 & 69.55 & 68.39 & 67.51\\
\hline
Env+I-attn & 69.85 & 67.00 & 68.40 & 68.72 \\
\hline
E-attn+Env & \textbf{70.10} & 72.39 & 71.23 & 70.45\\
\hline
\tabincell{c}{E-attn+Env+I-attn} & 69.91 & \textbf{75.45} & \textbf{72.58} & \textbf{71.19} \\
\hline
\end{tabular}
\end{table}

\subsection{Ablation Study}
In order to achieve a better understanding of the effectiveness of our proposed models, we perform ablation study and present the results in Table \ref{Experiment: ablation}. In Table \ref{Experiment: ablation}, E-attn is the abbreviation for explicit attention and I-attn is the abbreviation for implicit attention. Env indicates the model that concatenates user environment directly by a hierarchical structure. 

For our proposed models, since image and caption are fused in feature level, we use the Early Fusion model as our baseline. The ablation study results show that both E-attn and Env could achieve better improvement in accuracy and precision compared with Early Fusion. Besides, the performance of model improves as the number of additive structures increases. E-attn+Env model begins to surpass Early Fusion from all indexes. And finally, E-attn+Env+I-attn, namely Dual-attention model achieve best results among all combinations.

\begin{figure}
\centering
\includegraphics[width=3.5in]{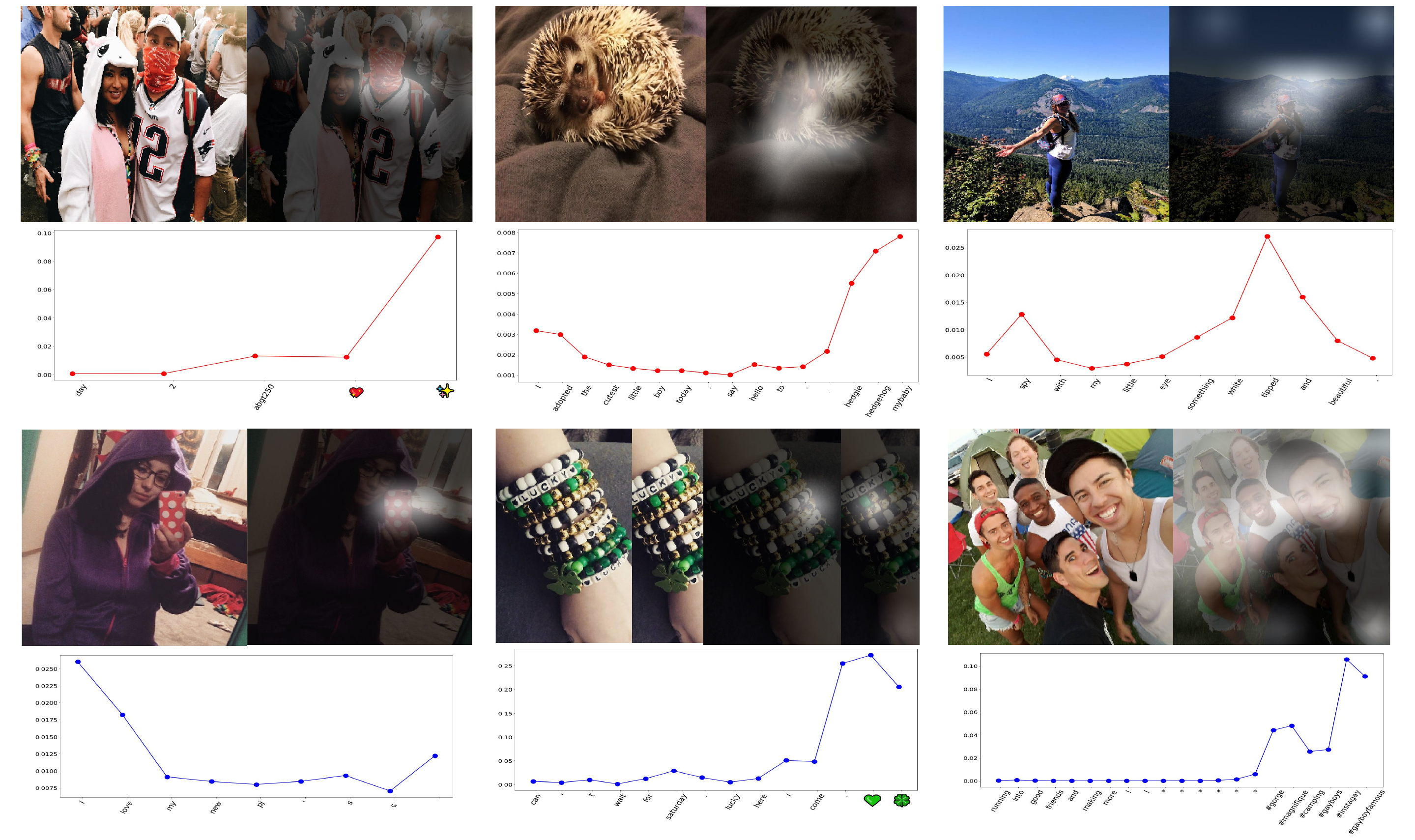}
\caption{ Visualization of the image attention maps and word attention weights. The first row corresponds to positive examples, where we demonstrate the word attention weights by the red line. The second row corresponds to negative examples, where we demonstrate the word attention weights by the blue line.}
\label{Fig: img_attn}
\end{figure}

\begin{figure}
\centering
\includegraphics[width=3.5in]{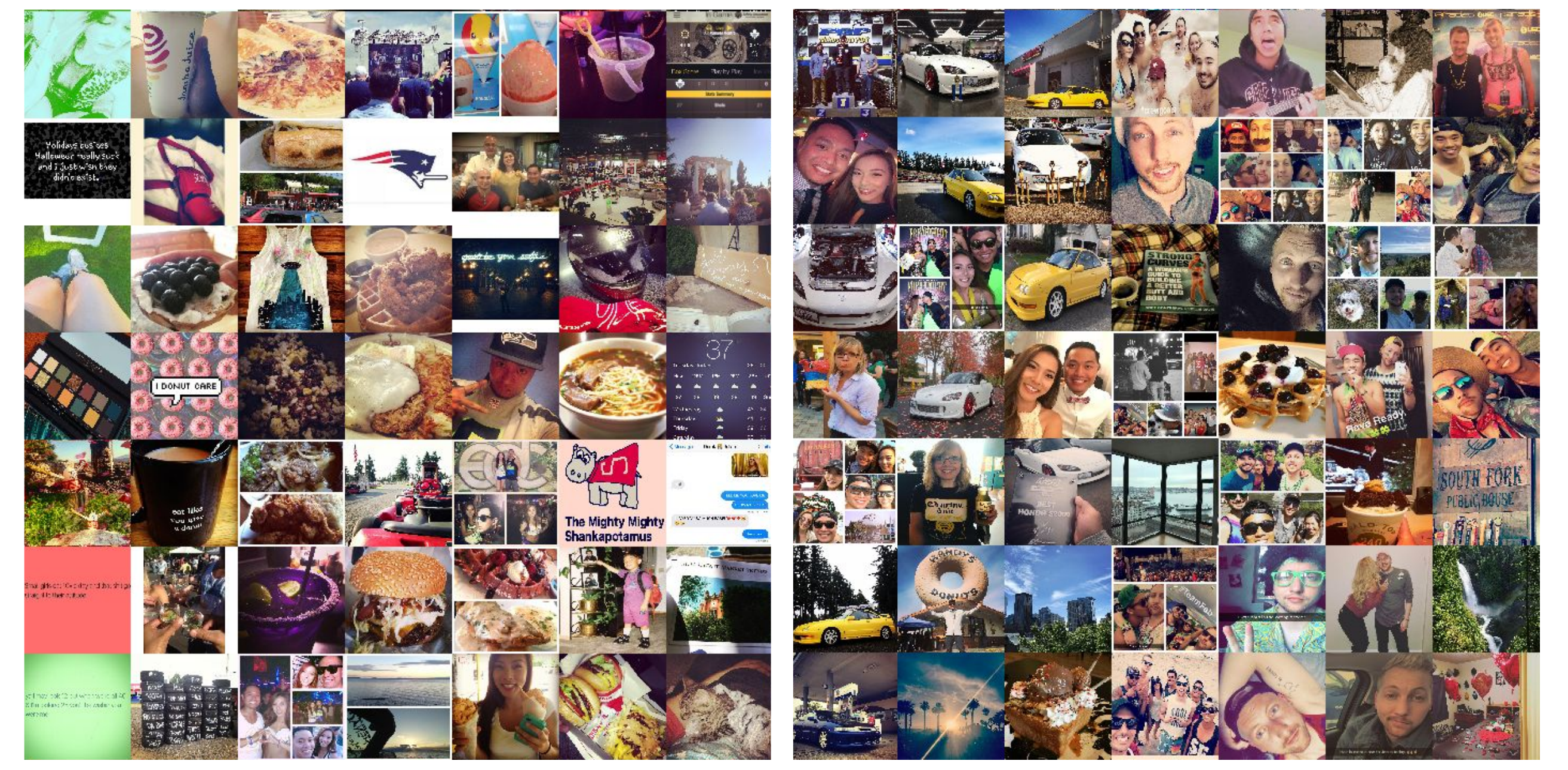}
\caption{ The clustering results for popular and unpopular images. Unpopular images are shown on the left hand side. Popular images are shown on the right hand side.}
\label{Fig: image_clustering}
\end{figure}

\begin{table}
\newcommand{\tabincell}[2]{\begin{tabular}{@{}#1@{}}#2\end{tabular}}
\caption{ Clustering categories and their corresponding ratio $R$}
\label{Table: kmeans}
\centering
\begin{tabular}{|c|c|c|}
\hline
Name & ratio $R$ \\
\hline
Photos of Daily Life & 0.475\\
\hline
Meal and Drink  & 0.696\\
\hline
Small Group Photo  & 0.396\\
\hline
Group Photo & 0.393\\
\hline
Poster& 0.760\\
\hline
Picture with a caption on& 0.627\\
\hline
Landscape Photo  & 0.602\\
\hline
Text Poster  & 0.700\\
\hline
Image combined by small pictures  & 0.492\\
\hline
Phone Screen Shot & 0.595\\
\hline
Selfie& 0.461\\
\hline
Car and Daily Commodities & 0.495\\
\hline
\end{tabular}
\end{table}

\subsection{Visualization}
To provide a deeper insight into what kind of things our model tends to focus on, we randomly select three pictures respectively from positive and negative samples, then visualize their image attention maps and word attention weights in Figure \ref{Fig: img_attn}. For image attention maps, we extract new image features from the last Convolutional layer of ResNet-50 first. The dimension of new image features is (7,7,2048). Unlike the original image features (2048 dimension vector), the new image features contain spatial information (7*7). We input the new image features into Dual-attention model part by part and generate a probability map for each image based on the probability score of each part. Therefore, different from the traditional image attention map, our image attention map actually reflect the popularity level of each zone. As shown in Figure \ref{Fig: img_attn}, human face, hedgehog, bracelet and mobile phone get relatively high popularity score compared with the other parts of the images. Based on this phenomenon, we conclude that concrete objects tend to get high popularity scores by our proposed model.

Under each image, we plot the word attention weights to illustrate the effectiveness of our explicit attention model. We can see that emoji, hashtag, specific object or specific action always tend to obtain high attention weights. For instance, in the first and fifth plots (from upper left to bottom right), the attention value of emoji ``star'', ``four-leaf clover'' and ``heart'' are much higher than the other words in the same captions. In the last plot, all hashtags obtain high attention scores. From the second and third plots, we can observe that attention weights of ``hedgehog'', ``hedgie'', ``spy'' and ``tipped'' increases remarkably, indicating that the model pays more attention to these words. Generally speaking, the explicit attention mechanism is able to capture keywords in captions and correlate them well with the image information.

\begin{figure}
\centering
\includegraphics[width=3.3in]{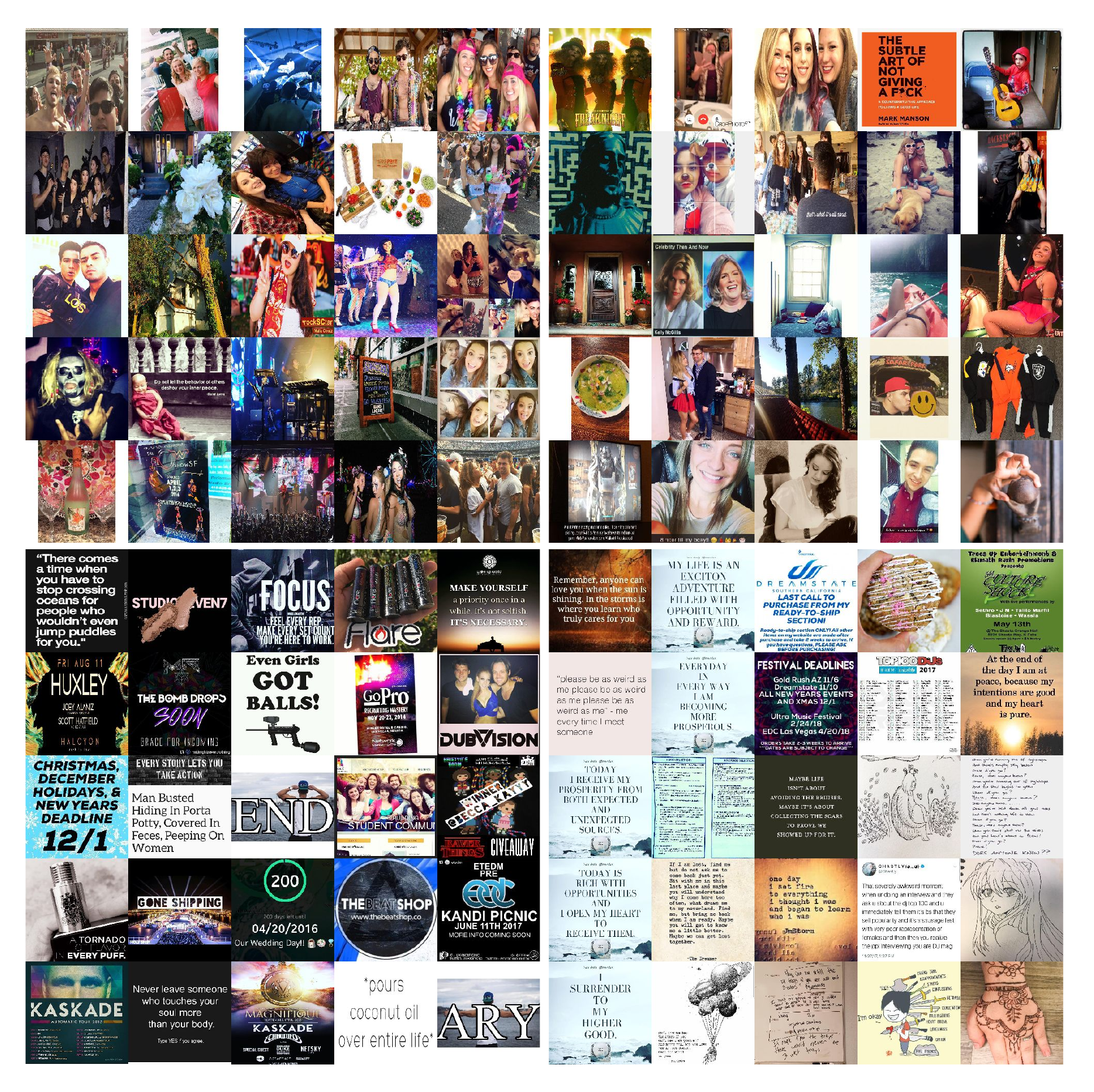}
\caption{ Selected categories by K-means clustering. From upper left to bottom right: 
Group Photo, Small Group Photo, Poster, Text Poster. }
\label{Fig: repeat_image_clustering}
\end{figure}

\subsection{Image Analysis}

\subsubsection{Intuitive Feeling}
\label{subsection: intuitive feeling}
Firstly, we would like to have an intuitive feeling about the difference between popular and unpopular images. As shown in Figure \ref{Fig: image_clustering}, popular images seem more complex and always contain objects like people, selfies, and so forth. On the other hand, unpopular images are simpler. Many of them are posters, advertisements, screenshots, or foods. 

\subsubsection{K-means Cluestering}

To get a more solid conclusion, we implement K-means Clustering on the high-level image features $V$ into 12 categories. For each category, the ratio of unpopular images to total images $R$ is set as an evaluation index. In order to obtain high-quality clustering results, we apply the following training strategy:

\begin{enumerate}[1.]
	\item Cluster the remaining images into $K$ classes and calculate the ratio $R$ of each class;
	\item Pick out the classes whose score $|R-0.5|$ is larger than threshold $t$;
	\item Repeat step 1 until there is no class satisfy the qualification in step 2;
    \item Collect the picked out and the remaining classes as the final results.
\end{enumerate}

By this strategy, the difference between popular and unpopular images of each category can be maximized. Table \ref{Table: kmeans} shows the clustering classes and their corresponding ratio $R$. We can see that categories such as ``poster'', ``meal and drink'' obtain high scores of $R$, which means images belong to these categories are likely to get a small number of ``like''. In contrast, classes like ``group photo'', ``small group photo'', and ``selfie'' get low scores of $R$, which means images in these categories tend to receive a large number of ``like''. This phenomenon is consistent with our intuitive conclusion in Section \ref{subsection: intuitive feeling}. To better understand our results, we select four typical categories according to their value of $R$ and demonstrate their pictures in Figure 7.

\subsection{Text Statistic Analysis}
Thanks to visualization, we already have an intuitive feeling about the connection between text and popularity. In this section, we analyze their correlation from a statistical perspective. We manually divide the captions into two categories: words and emojis. To avoid redundancy, we use ``text'' to represent the set of words and emojis. And for each category, we design the following experiment to reveal their correlations with popularity.

\begin{figure}
\centering
\noindent\makebox {
\begin{tabular}{c}
\includegraphics[width=0.98\linewidth]{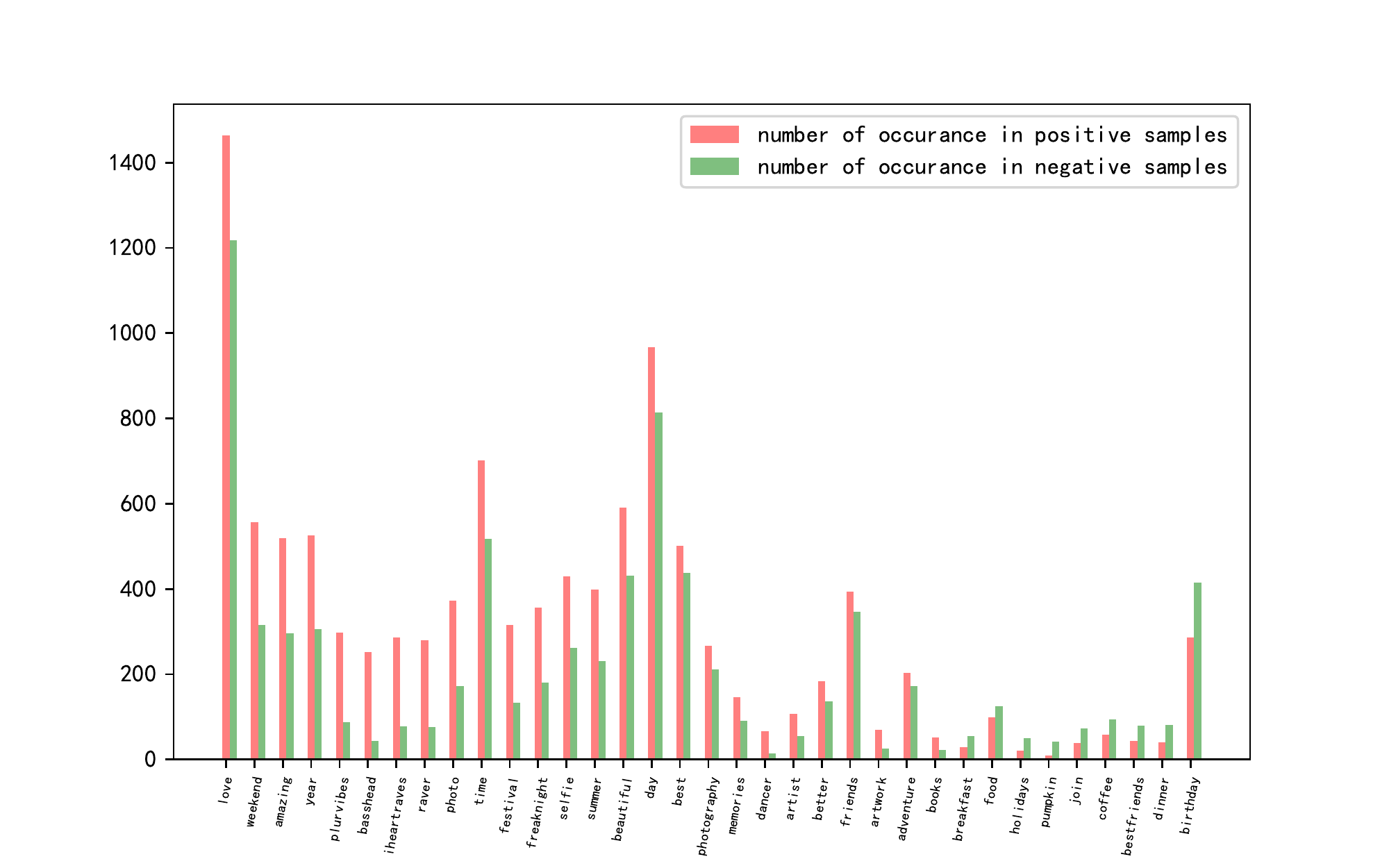} \\
(a)\\
\includegraphics[width=0.85\linewidth]{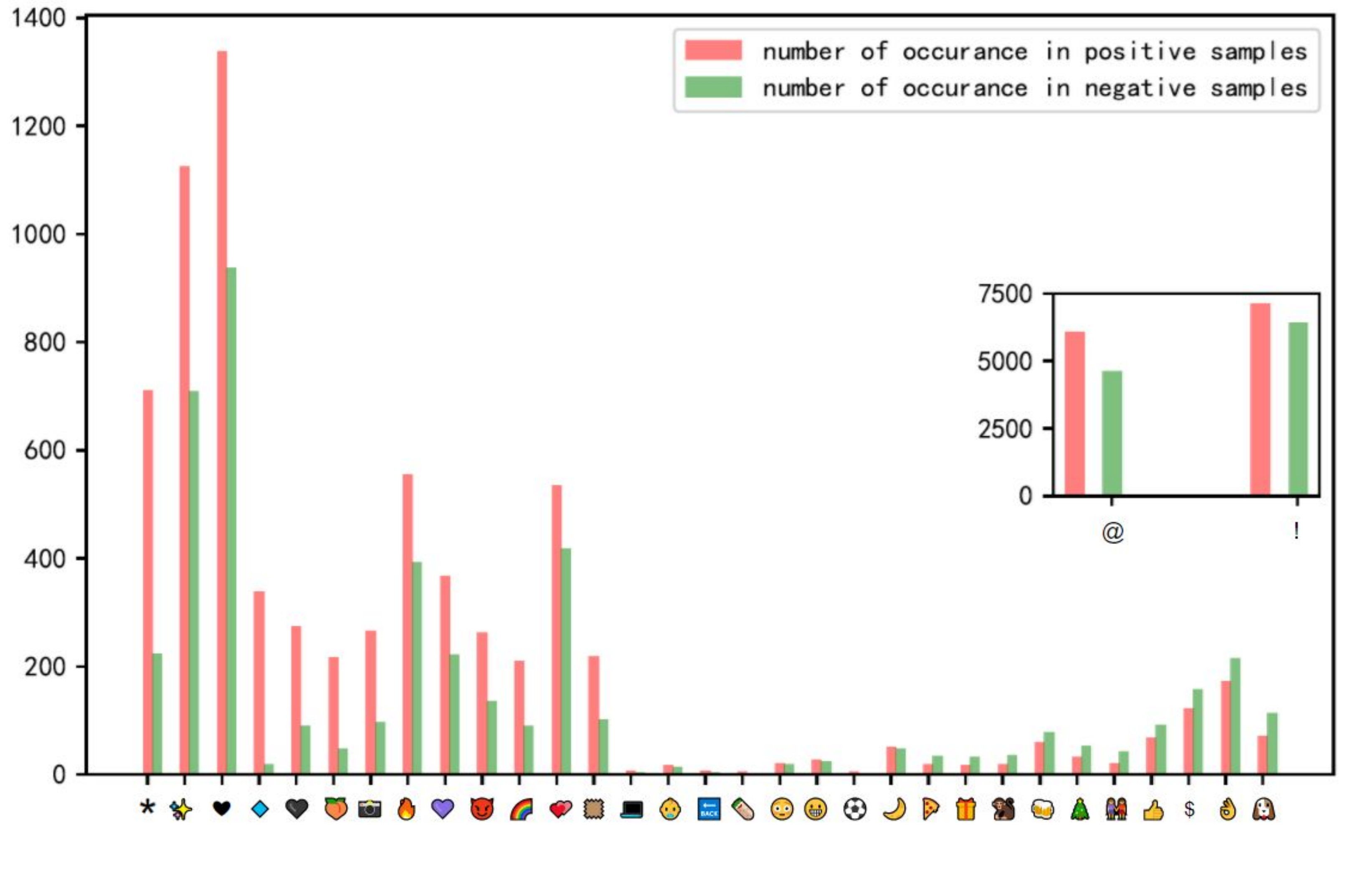}\\
(b)\\
\end{tabular}
} 
\caption{ The occurrence number of selected texts. (a) Statistic results of words, b) Statistic results of emojis. }
\label{Fig:occurrence_num}
\vspace{-0.3cm}
\end{figure}

In this experiment, we count the frequency of occurrence for each word. Take the word ``love'' as an example, we go through all the captions in positive samples and count the total occurrence number $m_p$ of ``love''. Then we perform a similar process to count the total occurrence number $m_n$ in negative samples. Since the ratio between positive and negative samples is 1:1, we do not need to normalize $m_p$ or $m_n$. In the end, the texts are ranked directly by the value of $m_p-m_n$. Before moving to the statistic analysis, we first filter out unrelated texts to avoid bias. These texts include basic symbols (``.'', ``$\cdot$'', ``:'', ...), pronouns (``this'', ``it'', ``that'', ...), prepositions (``with'', ``at'', ``of'', ...). There are some words that are used only in special conditions. For example, ``uscevents'', ``paradiso'' tend to appear in popular captions only from the users of USC (University of Southern California). On the other hand, ``chicago'', ``illinois'' always appear in popular posts from the users of UIUC (University of Illinois Urbana-Champaign). Hence we also delete this kind of words. After filtering out the unrelated texts, we demonstrate the occurrence number of typical texts in Figure \ref{Fig:occurrence_num}. We manually separate the words into three stages: top 25\%, median 50\%, and bottom 25\%. In contrast, we divide the emojis into two levels: top 50\% and bottom 50\%. The reason is that most emojis appear in popular posts and there is not a clear boundary between unpopular emojis and the other emojis. In Figure \ref{Fig:occurrence_num} (a), words from ``love'' to ``day'' are selected from the top 25\%. Based on statistic results, words that describe time (``year'', ``day'', ``time''), attribute (``amazing'', ``beautiful'') and correlated with holiday (``festival'', ``weekend'', ``selfie'') are very likely show up among the top 25\%. It seems that the posts include these words have higher tendency to receive ``like'' from other users. Words from ``best'' to ``books'' are selected from the median 50\%. We observe that most nouns distribute in this range for the reason that nouns are always used to describe objective events and they do not have much sentiment or subjective opinions involved. Words from ``breakfasts'' to ``birthday'' are selected from the bottom 25\%. We find that words describe food, such as ``coffee'', ``dinner'', ``breakfast'' appear frequently in this range, which is consistent with our image analysis results. Out of our expectation, ``holidays'', ``birthday'', ``bestfriends'' are in this region too. Just a guess, ``holidays'' is too general, it often appears in posters where people don't always give a ``like''. And when people mention ``birthday'', ``bestfriends'' in posts, only their close friends are likely to give them a ``like''. In Figure \ref{Fig:occurrence_num} (b), the first thirteenth emojis are selected from the top 50\% and the others are from the bottom 50\%. Since we do not find out an obvious regular variety among these emojis, we just list their statistic results and provide readers an intuitive feeling.

\section{Conclusions}
In this paper, we propose a dual-attention framework to address the image-caption based popularity prediction problem. Since our prediction target is for a specific user, we introduce the user environment as a high-level input to guide the classification model. User environment is incorporated by an implicit attention mechanism and image-caption pair is incorporated by an explicit attention mechanism. The classification results show that Dual-attention outperforms baselines on all measurements. Visualization results confirm that our model can clearly learn popularity words or image regions. Finally, we analysis image and textual information based on statistical results and draw conclusions about the correlation between image, caption, and popularity. In the future, we intend to develop a more efficient model to incorporate the user environment. Furthermore, a more comprehensive user profile by fusing locations, seasons, etc., can also be considered.

\bibliography{main}
\bibliographystyle{IEEEtran}

\end{document}